# Measurements of HCl and HF above Arizona from 1970 to 2012

## Lloyd Wallace[1] and William Livingston[2]

1 National Optical Astronomical Observatory, Tucson, AZ 85726

2 National Solar Observatory, Tucson, AZ 85726

Key points:

Stratospheric HCl and HF changes 1970 to 2012

Abstract.

We have observed at high resolution spectroscopic changes in Hydrogen Chloride (HCl) and Hydrogen Fluoride (HF) atmospheric trace gases from the 1970s through 2012. HCl rose in strength through 1998 and then decreased to 2012. HF has shown a continual increase to date. Both results are in general agreement with observations from the Jungfraujoch in the Swiss alps.

## 1) Introduction

HCl and HF are well-known as upper atmospheric trace gases, primarily the product of photo-dissociation of man-made halocarbons. With the reduction in the production of such halocarbons following the 1987 Montreal protocol the strong early increases in the amounts of HCl and HF have stopped. We have continued our measurements of these two gases in order to clarify their temporal behavior. The present work is directed to those objectives.

## 2) Observations

The long-term measurements require ground-based photometry of sharp absorption lines of the above gases superimposed on the solar spectrum. The best lines are those of fundamental vibration-rotation transitions in the infrared, which are all distorted to some degree by other absorptions of terrestrial origin. Most clear are the R(1) line of HCl at 2925.897 $cm^{-1}$ (3.4168 microns) and the R(1) line of HF at 4038.963 $cm^{-1}$ (2.475 microns).

Our observational material was all obtained with instruments at the McMath-Pierce solar telescope on Kitt Peak: altitude 2120 m, longitude -111 degrees, latitude 32 degrees. The spectra are limited by atmospheric transmission from 0.389 to about 20 microns. A number of spectra, actually taken for other purposes, contain HCl and HF lines. One such spectrum was taken by Don Hall in September 1971, with a 20 m focal length horizontal spectrograph, and this is the earliest HCl recording. Subsequently many useful spectra were obtained with the 1m Fourier Transform Sectrometer (FTS) [Brault, 1985]. During the interval between 1978 and 2005 observations of interest here were made by J. Brault, G. Stokes, D. Johnson, M. Brown, L. Delbouille, J.-M. Flaud, D. Deming and W. Livingston. The FTS did not function adequately after 2005 and Livingston began using the 13.5 m vertical spectrograph [Pierce, 1964] for HF in 1990. In 1992 a suitable infrared grating was installed which now accommodated both HCl and HF. Measurements with that system continues currently. The extraction of column amounts of the gases from these spectra is an extension of that given in Wallace, Livingston and Hall [1997].

## 3)  Results

Morning and afternoon averages of the column amounts determined from the three instruments cited above are given in the Figure. For HCl there are 139 data points from the FTS and 419 points from the spectrograph. For HF there are 400 points from the FTS and 181 from the spectrograph.

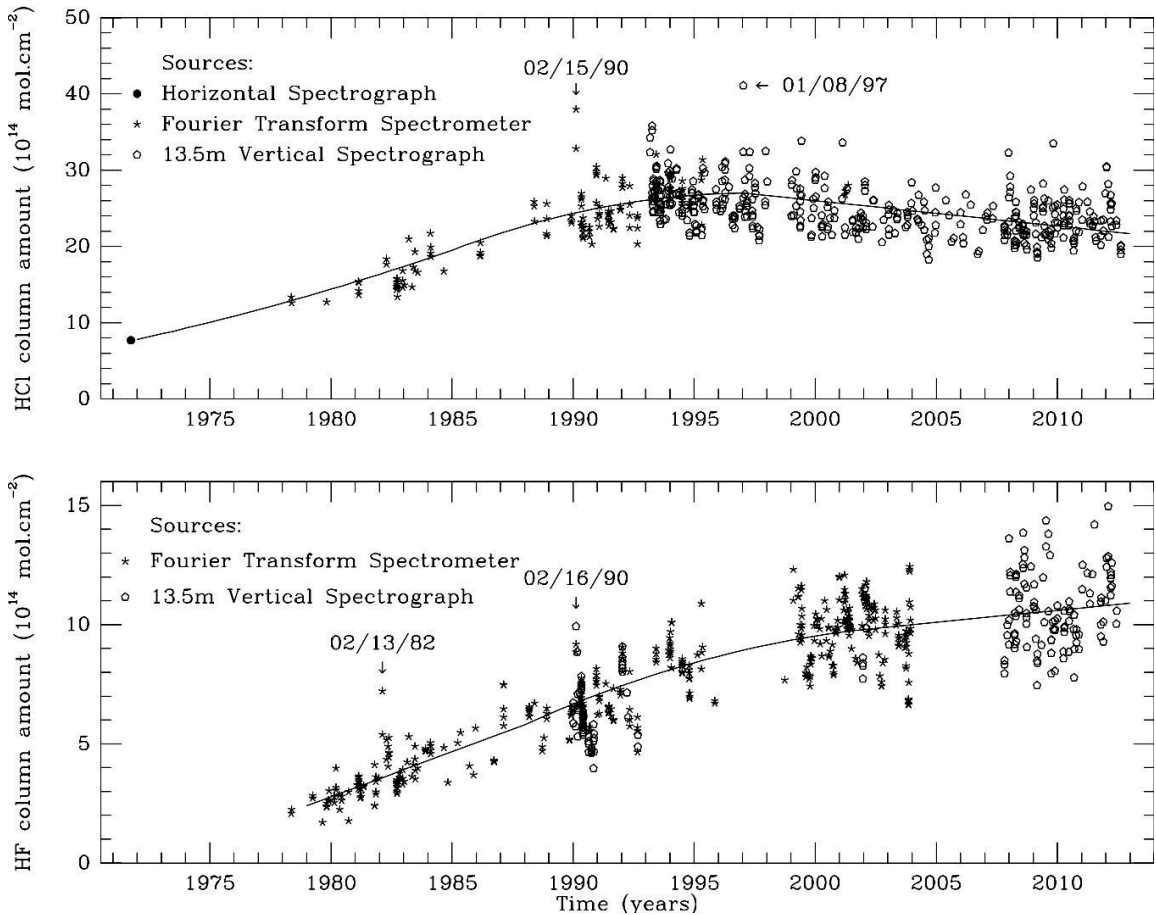

Figure. Top HCl, and bottom HF, temporal column amounts from the different instruments as indicated.

The substantial variation seen over short times are real; i.e. system noise is negligible at this scale [see Wallace et al. 1997]. The high points represent observations in January and February are represent seasonal effects as shown by the more frequent observations from the Jungfraujoch [Mahieu, 2004; Zander, unpublished]. We believe the latter may, however, be somewhat contaminated by European industrialization pollution. There is also the possibility of a contribution from volcanic emissions [Wallace and Livingston, 1992].

We have superimposed trend lines on the plots to emphasize our impression of the overall changes. For HCl we see a maximum in the 1993-1998 period followed by a slow decrease. Rinsland et al [2003] have used an entirely different set of Kitt Peak FTS spectra from 1977 to the end of 2001 and they found a peak in HCl in 1995 with an uncertainty of two years. Our data is more time dense over this period than Rinsland's and extends to mid 2012. As mentioned, data from the University of Liege FTS on Jungfraujoch gives an HCL maximum in about 1997 with a slow decrease following and a possible second maximum in 2004.

We see general increases in HF through the entire period as have Rinsland et al [2002], Zander (unpublished) and Duchatelet et al [2010].

## 4) *Conclusions*

It does indeed appear that HCl over the United States maximized in the 1993-1998 period and is now in slow decline. HF has not reached a maximum but its growth slowed after 2000 to almost no increase.

# References

Brault, J.W. (1985), Fourier transform spectroscopy, in High Resolution Astronomy, A. Benz, M. Huber, and M. Mayer, Eds., Swiss Society of Astrophysics and Astronomy, Geneva Observatory, 1-61.

Duchatelet, P., et al. (2010), Hydrogen fluoride total and partial column time series above the Jungfraujoch from long-term measurements., J. Geophys. Res., 115, D22306.

Hall, D.N.B (1970), Observation of the infrared sunspot spectrum between 11340 A and 24778 A, PhD thesis, Harvard University.

Mahieu, E., et al. (2004), The evolution of inorganic chlorine above the Jungfraujoch Station: an update, Ozone, 2, Proc. Of the XX Quad. Ozone Symp, Kos, Greece, 997-998.

Pierce, A.K. (1964), The McMath solar telescope of the Kitt Peak National Observatory, Appl. Opt. 3, 1337.

Rinsland, C.P., et al. (1991), Infrared measurements of HF and HCl total column abundances above Kitt Peak, 1977-1990: seasonal cycles, long-term increases and comparisons with model calculations, J. Geophys. Res. 96D, 15,523-15,540.

Rinsland, C.P., et al. (2002), Stratospheric HF column abundances above Kitt Peak (31.9 deg N latitude): trends from 1977 to 2001 and correlations with stratospheric HCl columns, J. Quant. Spectros. Rad. Transfer, 74,205-216.

Rinsland, C.P., et al. (2003), Long-term trends of inorganic chlorine from ground-based solar spectra: Past increases and evidence for stabilization, J. Geophys. Res., 108, 4252.

Wallace, L.W. and Livingston (1992), The effect of the Pinatubo cloud on hydrogen chloride and hydrogen fluoride, Geophys. Res. Lett. 19, 1209.

Wallace, L.W. Livingston, W., and Hall, (1997), A twenty-five year record of stratospheric hydrogen chloride, Geophys. Res. Lett. 24, 2363-2366.